\documentclass[useAMS,usenatbib,usegraphicx]{mn2e}
\usepackage{epsf, epsfig}
\usepackage{Times}

\def\be{\begin{equation}}
\def\ee{\end{equation}}

\def\be{\begin{equation}}
\def\ee{\end{equation}}
\catcode`\@=11 
\def\@versim#1#2{\vcenter{\offinterlineskip
\ialign{$\m@th#1\hfil##\hfil$\crcr#2\crcr\sim\crcr } }}
\def\lsim{\mathrel{\mathpalette\@versim<}}
\def\gsim{\mathrel{\mathpalette\@versim>}}

\newcommand{\arcdeg}{\ensuremath{^{\circ}}}

\title[X-rays from Quiescent Black Holes]{Origin of X-ray Emission from 
Transient Black Hole Candidates in Quiescence}
\author[Pszota et al.]
{Gabor Pszota$^{1}$\thanks{E-mail: pszotag@physics.purdue.edu}, 
Hui Zhang$^{2}$\thanks{E-mail: hzhang@shao.ac.cn}, 
Feng Yuan$^{2,3}$\thanks{E-mail: fyuan@shao.ac.cn},
Wei Cui$^{1}$\thanks{E-mail: cui@physics.purdue.edu}
\\
$^{1}$\/Department of Physics, Purdue University, West Lafayette, IN 47907\\
$^{2}$\/Shanghai Astronomical Observatory, Chinese Academy of Sciences, 
200030 Shanghai \\
$^{3}$\/Joint Institute for Galaxy and Cosmology of Shanghai Astronomical 
Observatory and University of Science and Technology of China}

\begin{document}

\date{Received 2008}

\pagerange{\pageref{firstpage}--\pageref{lastpage}} \pubyear{2008}

\maketitle

\label{firstpage}

\begin{abstract}
We report results from a systematic study of X-ray emission from black hole 
transients in quiescence. In this state mass accretion is thought to follow 
the geometry of an outer optically thick, geometrically thin disc and an 
inner optically thin, geometrically thick radiatively inefficient accretion 
flow (RIAF). The inner flow is likely also coupled to the jets near the
black hole that are often seen in such systems. The goal of the study is to 
see whether the X-ray emission in the quiescent state is mainly powered by 
the accretion flow or the jets. Using data from deep {\it XMM-Newton} 
observations of selected black hole transients, we have found that the 
quiescent X-ray spectra are, to a high precision, of power-law shape in 
the cases of GRO~J1655$-$40 and V404~Cyg. Such spectra deviate significantly 
from the expected X-ray spectrum of the RIAF at very low accretion rates. 
On the other hand, they can naturally be explained by emission from the 
jets, if the emitting electrons follow a power-law spectral distribution 
(as is often assumed). The situation remains ambiguous in the case of 
XTE~J1550$-$564, due to the relatively poorer quality of the data. We discuss 
the implication of the results.
\end{abstract}


\section{Introduction}

The majority of X-ray binaries that are known to contain a stellar-mass black 
hole are transient X-ray sources. They spend most of their time in the
quiescent
state, in which the mass accretion rate is thought to be extremely low. 
Occasionally, they undergo an outburst during which they may become the 
brightest X-ray sources in the sky. The exact mechanism that triggers such
an outburst is not entirely understood but is thought to be related to a 
sudden surge in the accretion rate that is caused by a thermal instability 
in the accretion disc (see reviews, e.g., by King 1995 and Lasota 2001). 
During an outburst, the X-ray properties of a black hole transient is often 
described empirically in terms of spectral states (e.g., McClintock \& 
Remillard 2006; Xue, Wu, \& Cui 2008). 

It is proposed that the spectral states may correspond to different 
configurations of the underlying accretion process at different mass 
accretion rates (Narayan 1996; Narayan, McClintock \& Yi 1996;
Esin, McClintock, \& Narayan 1997). Specifically, when the accretion rate 
is high in the high-soft state, the accretion flow is thought to follow the 
geometry of the Shakura \& Sunyaev (1973) disk (SSD), which is geometrically 
thin and optically thick. This can naturally explain the observed 
blackbody-like X-ray spectrum that is characteristic of the high-soft state. 
As the accretion rate decreases, the source evolves towards the low-hard 
state. In the process, a phase transition is thought to occur in the inner 
portion of the disc, in which the accreted matter is heated to nearly local 
virial temperatures and may also form an outflow wind (see Narayan 2005 for 
a comprehensive review of the models, their evolution, and their applications 
to black hole candidates and active galactic nuclei). The accretion flows in
this region is a geometrically thick but optically thin configuration, and 
is radiatively inefficient. The accretion power is mostly advected into the 
black hole or carried away by the outflow (Blandford \& Begelman 1999). Such 
a radiatively inefficient accretion flow (RIAF) is 
capable of producing hard X-rays by up-scattering ambient soft photons. This 
can naturally explain the increasing dominance of the power-law component 
of the X-ray spectrum, as the source approaches the low-hard state. The RIAF 
model predicts that the trend continues towards even lower accretion rates, 
as more of the accretion flow becomes advection dominated, and that the 
X-rays in the quiescent state originate entirely from the RIAF (Narayan, 
McClintock, \& Yi 1996). 

A more recent development is the realization of the potentially critical role
of jets, which seem to be ubiquitous in black hole transients (see review by 
Fender 2006 and references therein). Yuan, Cui \& Narayan (2005) demonstrated 
that it would be nearly impossible for the RIAF model, which is quite 
successful in explaining the X-ray emission from black hole transients in
the low-hard state, to 
also account for the observed emission at longer wavelengths (radio and IR 
in particular). In order to describe the broadband spectral energy 
distribution (SED) in the low-hard state, they showed that contributions 
from both accretion flow and jets would be needed, with the former 
mainly responsible for emission at UV/X-ray wavelengths, the latter for 
emission at radio/IR wavelengths, and both for emission in between (cf. 
Malzac, Merloni, \& Fabian 2004, who argued that the optical emission might 
be dominated by the jets), when 
the accretion rates are relatively high. Extrapolating the result of Yuan et 
al. (2005) to lower accretion rates, Yuan \& Cui (2005) predicted that the 
X-ray emission from the jets would eventually exceed that from the hot flow 
(because the former is proportional to the accretion rate $\dot{m}$, 
normalized to the Eddington rate, and the latter roughly to $\dot{m}^2$). 
In the quiescent state, the X-ray emission should, therefore, be mainly 
powered by the jets, at variance with the prediction of the RIAF model. 
Falcke, Kording \& Markoff (2004) also postulated that the X-ray emission of 
the quiescent state would be dominated by the jets. But, in contrast to 
Yuan \& Cui (2005), they argued that this would be the case even for the 
low-hard state.

Is it the accretion flow or jets that power the X-ray emission from black hole 
transients in the quiescent state? Yuan \& Cui (2005) proposed two 
observational tests to answer the question. If the quiescent X-rays are 
powered by the jets, they predicted: (1) the radio/X-ray correlation 
would steepen when the X-ray flux drops below a characteristic value and 
(2) the X-ray spectrum would be of power-law shape. It has been claimed 
that the observation of A0620-00 is at odds with the first prediction (Gallo 
et al. 2006). However, the conclusion hinged critically on a radio/X-ray 
correlation that had been thought to hold for all black hole candidates. 
The universality of the radio/X-ray correlation has since been brought into
question (Xue \& Cui 2007). The second prediction 
is a viable test because the X-ray spectrum of an RIAF would deviate 
strongly from power-law shape at sufficiently low accretion rates, when the 
density of the flows becomes so low that Comptonization is dominated by 
single scattering (Narayan, McClintock, \& Yi 1996; Quataert \& Narayan 1999; 
McClintock et al. 2003; Yuan, Cui, \& Narayan 2005).

In this work, we present results from a systematic study of black hole 
transients in quiescence. A number of such sources had been observed and 
detected 
earlier with {\it Chandra} and {\it XMM-Newton} (Kong et al. 2002; Hameury 
et al. 2003) but none of the X-ray spectra obtained are of sufficiently 
high quality that would allow us to distinguish jet-based and accretion-based 
models. To improve the situation, we carried out deep observations of 
selected sources with {\it XMM-Newton}. The results reported here are based 
on data from these as well as an archival {\it XMM-Newton} observation.

\section{Observations and Data Analysis}

Based on information on the quiescent X-ray fluxes of transient black holes
from previous works (Kong et al. 2002; Hameury et al. 2003), we selected
V404~Cyg, XTE~J1550$-$564, and GRO~J1655$-$40 for this pilot study. V404~Cyg
is the brightest in X-rays among all black hole transients that have been 
observed in the quiescent state and is thus the best source for our 
investigation. It had already been observed with {\it XMM-Newton} for about
40 ks, with the EPIC detectors in the full-window mode and the medium filter
used. We estimated that the quality of the data from that observation would
be sufficient for our purposes. For this work, therefore, we simply used the
archival data.

XTE~J1550$-$564 had never been observed with {\em XMM-Newton} before, despite 
being the first black hole candidate to have its jets directly imaged in 
X-rays and also being one of the most active, producing an outburst (of 
varying magnitude) every couple of years. The quiescent X-ray luminosity 
of XTE~J1550$-$564 is nearly the same as that of V404~Cyg ($\sim 10^{33}$ 
$ergs$ $s^{-1}$; Kong et al. 2002) but it lies at a larger distance.
It seems to have a harder X-ray spectrum than V404~Cyg but the uncertainty 
is quite large (Kong et al. 2002). We observed XTE~J1550$-$564 in quiescence 
with {\em XMM-Newton} for about 60 ks in 2007, with the EPIC detectors also 
in the full-window mode but the thin filter was used. Unfortunately, the 
observation suffered from severe contamination by solar flares. The standard 
filtering procedure (as described in the {\it XMM-Newton} Data Analysis 
Cookbook\footnote{See http://wave.xray.mpe.mpg.de/xmm/cookbook}) removed 
the large flares but smaller ones were still visible after the filtering. 
We performed additional filtering, based on the 0.5--10 keV lightcurve of 
the whole chip, to minimize the effects of the contamination. The resulted 
effective exposure times are only about 
14~ks and 24~ks for the pn and MOS detectors, respectively.

GRO~J1655$-$40 is also relatively luminous in X-rays in the quiescent state
but the luminosity seems to vary greatly, from about $3\times 10^{31}$ to 
$10^{33}$ ergs/s (Hameury et al. 2003). The source is at roughly the same
distance as V404~Cyg. It had already been observed by {\em XMM-Newton} for
about
50 ks (Hameury et al. 2003). The X-ray flux measured is about a factor of 5 
less than that of XTE~J1550$-$564. Since a similar flux had been obtained 
earlier with {\em Chandra} (Kong et al. 2002), this is perhaps the more
typical state for the source. The X-ray spectrum of GRO~J1655$-$40 is quite 
similar to that of XTE~J1550$-$564 but the spectrum lacks statistics for 
our purposes. We observed GRO~J1655$-$40 in quiescence with {\it XMM-Newton} 
in 2007 for about 185~ks. The EPIC detectors were operated in the 
full-window mode with the thin filter. We should note that this observation 
was also partially contaminated by solar flares, but not as severely as in 
the case of XTE~J1550-564.

Table~1 summarizes the key characteristics of the {\em XMM-Newton} 
observations. We followed the same procedures, as described in the Data 
Analysis Cookbook, for all observations in preparing and filtering data, 
making light curves, extracting spectra, and generating the corresponding 
response matrix files and ancillary response files for subsequent spectral 
modeling. Briefly, the source counts were extracted from within a circular 
region that is centered on each source and is of radius 40\arcsec\ 
(or roughly the 85\% encircled-energy radius of the on-axis PSF), which is 
the largest source region that could be used without extending to a different 
CCD chip. The background counts were obtained from a different circular 
region of radius 80\arcsec\ that is located on the same chip, sufficiently
far from any sources.

The raw pn and MOS spectra were re-binned such that each energy bin contains
at least 16 counts. They were then filtered to cover the energy range of
0.5--10~keV. A 1\% systematic error was added to the spectra. The initial
modeling was carried out in {\it XSPEC} (Arnaud 1996). We chose to jointly
fit the individual pn and MOS spectra. For XTE~J1550$-$564, however, we 
only used the pn spectrum, because of the poorer statistics of the MOS data, 
and limited the energy range to 0.5--7.7~keV, where the signal-to-noise 
ratio is adequate. In all three cases, the X-ray spectrum can be fitted 
satisfactorily with a simple power law that is attenuated by interstellar 
absorption. Figure~1 shows the X-ray spectra of the three sources, along
with the best-fit power laws and residuals. The model parameters are 
summarized in Table~2. Note that the hydrogen column density was fixed 
to the line-of-sight value in the case of XTE~J1550$-$564; the actual 
value should be lower but it would hardly affect our results, again due to 
the poor quality of the data. Our results on V404~Cyg are in agreement with 
those previously obtained from the same dataset (Bradley et al. 2007).

In each case, we used the best-fit power law to compute unabsorbed photon 
and energy fluxes for a set of energy bins, to quantify the intrinsic
X-ray spectrum of each source. The widths of the bins (typically 
0.2$-$0.3~keV) were chosen to maintain sufficiently small error bars on
the fluxes across the entire energy range. The error bars were derived (in 
XSPEC) by randomly drawing 1000 sets of parameters and taking the $1~\sigma$ 
range of the resulted flux distribution. From the photon and energy fluxes 
we also computed the effective energy of each bin ($E_{eff}=F_{en}/F_{ph}$). 
The unabsorbed X-ray spectra are shown in Figure~2. It is quite apparent 
that the quiescent spectra of GRO~J1655$-$40 and V404~Cyg are, to a high
precision, of power-law shape. The situation is a bit ambiguous in the case 
of XTE~J1550$-$564, due to the limited statistics of the data.

\section{Modeling Results}

We proceeded to examine the derived spectra with the same coupled 
accretion$-$jet model that has provided a good description of the SED of 
XTE~J1118+480 over a broad spectral range (from radio to hard X-ray 
frequencies; Yuan, Cui, \& Narayan 2005). In this model, the accretion flow 
is described as a standard thin disk outside a transition 
radius $r_{\rm tr}$ and an RIAF inside this radius. For the quiescent state, 
$r_{tr}$ is expected to be very large (Narayan 2005), hence the disc is 
cool and likely contributes only to IR/optical emission. The effect of 
outflow/convection in the RIAF is taken into account in calculating the 
dynamics of the RIAF. The main parameters of RIAF are the viscous parameter 
$\alpha$, a parameter describing the strength of the magnetic field $\beta$, 
and a parameter $\delta$ which determines the fraction of the turbulent 
dissipation which directly heats electrons. For the radiation processes 
in the RIAF, we considered synchrotron emission, bremsstrahlung emission,
and inverse Compton scattering. Near the black hole, we assumed that 
a fraction of the accretion flow is transferred into the vertical 
direction to form a jet. Within the jet, internal shocks occur 
due to the collision of shells with different velocities. These shocks 
accelerate a fraction of the electrons into a power-law energy distribution. 
The steady-state energy distribution of the accelerated electrons 
is self-consistently determined, taking into account the effect 
of radiative cooling. The energy density of accelerated electrons 
and amplified magnetic field is determined by two parameters, $\epsilon_e$ and 
$\epsilon_B$, which describe the fraction of the shock energy going 
into electrons and the magnetic field, respectively. Then the synchrotron 
radiation from these accelerated electrons can be calculated. We note that 
the effects of inverse Compton scattering of the synchroton photons in the 
jet are negligible due to small scattering optical depth. We refer the 
readers to Yuan et al. (2005) for a detailed description of the model.

For this work, we fixed all model parameters to the values derived from 
fitting the SED of XTE~J1118+480 (Yuan, Cui, \& Narayan 2005), except 
for those that are source specific, such as distance, inclination angle, 
black hole mass, accretion rate, and $r_{tr}$, which is presumed to depend 
on accretion rate. As before, we assumed that the accretion rate is 
dependent of radius in the RIAF, $\dot{M}=\dot{M}_{0}(r/r_{tr})^{0.3}$ 
(Yuan, Cui, \& Narayan 2005). However, we did {\em not} assume that the 
accretion rate is constant in the optically thick disc, because the disc 
is not expected to be in a steady state for the quiescent state (e.g., 
Lasota 2001). Instead, the accretion rate is expected to decrease with 
radius, resulting in a roughly isothermal temperature profile, which 
deviates significantly 
from the SSD. For this work, therefore, we adopted a constant-temperature 
disc for the model (although we also investigated scenarios in which the
disc is assumed to be of SSD type). The results of this modeling are also 
shown in Fig.~2, with the model parameters summarized in Table~3. 

We also explored the possibility of modeling the X-ray spectra with a pure
accretion model (consisting only of disc plus RIAF). The results from our 
best attempts are also shown in Fig.~2 (insets); the values of the model 
parameters are shown in Table~3. We should note that in the cases of 
XTE~J1550-564 and GRO~J1655-40 we had to adopt a lower value for the 
viscous parameter in the RIAF (not shown in the table), $\alpha=0.1$ as 
opposed to $0.3$ (see Yuan, Cui, \& Narayan 2005), to achieve better fits 
to the data.
As is apparent from Fig.~2, the high-quality quiescent X-ray spectra of 
GRO~J1655$-$40 and V404~Cyg cannot be accounted for by emission from the 
RIAF alone. We believe that this conclusion is robust, independent of 
theoretical uncertainties associated with the model adopted. Fundamentally, 
at very low accretion rates, contribution from multiple scatterings in 
the hot flows is negligible, so the Comptonized spectrum deviates strongly
from a power law. On the other hand, the observed power-law shape of the 
quiescent X-ray spectra can be naturally explained by emission from the 
jets, if the spectral energy distribution of the emitting electrons is, 
as often assumed, of power-law shape. Therefore, our results are in favor 
of a jet origin of the X-ray emission from black hole transients in 
quiescence, lending support to the prediction of Yuan \& Cui (2005).

The situation is still ambiguous in the case of XTE~J1550$-$564, due to the 
relatively large uncertainties of the measurements. An RIAF origin of the 
X-ray emission cannot be ruled out in this case (see Fig.~2). A deeper 
{\em XMM-Newton} observation of the source 
in quiescence would provide the much needed data to resolve the ambiguity. 
In general, we believe that it is important to extend the effort to more
 black hole transients. The data would allow us to see whether the 
conclusion holds for such systems as a population.

For comparison, we also included in Fig.~2 the published results on the 
quiescent IR/optical fluxes of the disc (Casares et al. 1993; Orosz et al. 
2002; Greene et al. 2001). It should, however, be stressed that these 
measurements were {\em not} made simultaneously with the X-ray measurements 
and should thus be taken with caution, because the sources can vary 
significantly even in the quiescent state (e.g., Bradley et al. 2007). 
Extrapolating our ``best-fit'' models to the IR/optical wavelength, we 
found that the predicted fluxes fall below the values from earlier
measurements for XTE~J1550$-$564 and V404~Cyg. We found that the discrepancy
was worse when we replaced the constant-temperature disc with the SSD; the
real disc is likely to fall in between these two extreme scenarios. The
discrepancy could be attributed
to the variability of the sources in the quiescent state. For instance, the 
accretion rates might have been higher during the IR/optical observations, 
so the discs were brighter then. Alternatively, there might be additional 
sources of IR/optical emission, for instance, hot spots where gas from the 
companion star impacts the accretion disc (cf. McClintock et al. 2003). 
We should note, however, that in both cases we were able to find good fits 
also to the IR/optical data by adopting different values for the distance 
and inclination angle of the systems (that are still within measurement 
uncertainties). 

\section{Discussion}

Fender et al. (2003) argued, on the basis of the ``universal radio/X-ray 
correlation'', that the energetics of the quiescent state ought to be 
dominated by the jets, in the sense that the kinetic power of the jet is 
much greater than the X-ray luminosity of the accretion flows. In their 
jet-dominated state, however, the X-ray luminosity of the jet is not 
necessarily also greater than that of the accretion flows, because the 
radiative power of the jet is only of the order of 1\% of the kinetic 
power (see Yuan \& Cui 2005 for a more detailed discussion). Here, we have 
shown that the quiescent X-rays from transient black hole candidates are 
likely to originate from the jets, as opposed to the accretion flows.

In summary, we have, for the first time, found direct evidence that the 
quiescent state may be fundamentally different from the low-hard state, 
as far as the source of X-ray emission is concerned. Contrary to the view 
that the former may be a simple extension of the latter towards lower 
accretion rates, our results suggest that the X-ray emission from transient 
black holes is dominated by contribution from the jets (or other sources
of non-thermal electrons) in the quiescent
state, while in the low-hard state it is likely dominated by contribution 
from the RIAF (e.g., Esin, McClintock, \& Narayan 1997; Yuan et al. 2007). 

Finally, we would like to emphasize that the prediction of Yuan \& Cui (2005) 
is insensitive to the mass of black holes and might thus also hold for 
active galactic nuclei (AGN). Wu, Yuan \& Cao (2007) has recently modeled a 
sample composed of eight FR I galaxies and found that their X-ray spectra
should be dominated by jets, rather than by RIAFs, if their luminosities are 
below $\sim 10^{-6}L_{\rm Edd}$, and vice versa, as predicted by Yuan \& Cui 
(2005). Wrobel, Terashima, \& Ho (2008) observed two low-luminosity AGN at 
8.5 GHz and found that the observed radio luminosity is within a factor of 
3 of the value that is predicted from the observed X-ray luminosity and the 
radio--X-ray--mass relation derived by Yuan \& Cui (2005). 

\section*{Acknowledgments}
This research has made use of data obtained through the High Energy 
Astrophysics Science Archive Research Center Online Service, provided by 
the NASA/Goddard Space Flight Center. The work was supported in part by 
NASA through grants NNX07AQ29G and NNX07AH43G and by Natural Science 
Foundation of China 
through grant 10773024. F.Y. also acknowledges support from the Bairen 
Program of Chinese Academy of Sciences.

\clearpage

\begin{table}
\caption{Summary of Observations}
\begin{tabular}{ccccccc} \hline \hline
Obs. ID & Source & Date & Duration & \multicolumn{3}{c}{Effective Exposure}  \\
& & & (ks) & \multicolumn{3}{c}{(ks)} \\
\cline{5-7} \\
& & & & pn & MOS1 & MOS2 \\ \hline
0400890201 & GRO~J1655$-$40 & 2007.03.26 & 120 & 54.1 & 75.3 & 73.6  \\
0400890301 & GRO~J1655$-$40 & 2007.02.22 & 65 & 48.8 & 55.3 & 54.0 \\
0400890101 & XTE~J1550$-$564 & 2007.02.25 & 59 & 13.6 & 24.4 & 23.7 \\
0304000201 & V404~Cyg & 2005.11.08 & 40 & 31.0 & 36.8 & 36.4  \\ \hline
\end{tabular}
\end{table}

\begin{table}
\caption{Results of Power-Law Fitting$^1$}
\begin{tabular}{lcccc} \hline \hline
Source & $N_H$ & $\Gamma$ & $F_{x}$ & $\chi^{2}_{\nu}/dof$ \\
& $(10^{21} cm^{-2})$ & & $(10^{-14} erg s^{-1} cm^{-2})$$^b$ & \\ \hline
GRO~J1655$-$40 & $6.8^{+1.4}_{-1.1}$ & $2.0^{+0.1}_{-0.2}$ & $5.8^{+0.8}_{-1.8}$ & $1.06/374$ \\
XTE~J1550$-$564 & $9$ (fixed)$^a$& $2.1^{+0.6}_{-0.5}$ & $4.9^{+2.1}_{-4.6}$ & $0.96/20$ \\
V404~Cyg & $9.4^{+0.8}_{-0.7}$ & $2.11\pm0.09$ & $40^{+4}_{-5}$ & $0.90/285$ \\
\hline
\end{tabular}
\\$^1$ The errors shown represent 90\% confidence intervals for a single 
parameter. \\
$^a$ Dicky \& Lockman 1990. \\
$^b$ Measured X-ray flux in the 0.5--10~keV band. 
\end{table}

\begin{table}
\caption{RIAF+jet vs. pure RIAF modeling$^1$}
\begin{tabular}{lccccccccc} \hline \hline
Source&$d^b$&$M_{BH}^b$&$L_{x}/L_{Edd}$&$i^b$&$r_{tr}$&$\dot{m}(r_{tr})$&$\theta_{jet}$&$\dot{m}_{jet}$&$L_{kin}/\dot{M}(10R_{S})c^2$ \\
&(kpc)&($M_{\sun}$)&$(10^{-7})$&(\arcdeg)&$(R_S)$&$(10^{-4})$&(\arcdeg)&$(10^{-6})$&$(10^{-3})$ \\ \hline
GRO~J1655$-$40 & 3.2 & 6.3 & 1.4 & $70$ & $5000$ & $1.3$ & $85^c$ & $2.35$ & $28$ \\
XTE~J1550$-$564 & 5.3 & 9.6 & 2.3 & $73$ & $5000$ & $0.6$ & $73^d$ & $2.4$ & $62$ \\
V404~Cyg & 3.5 & 11.7 & 7.0 & $56$ & $5000$ & $5.0$ & $56^d$ & $2.8$ & $8.6$ \\ 
\hline
GRO~J1655$-$40$^a$ & 3.2 & 6.3 & 1.4 & 70 & 1000 & 1.5 & - & - & - \\
XTE~J1550$-$564$^a$ & 5.3 & 9.6 & 2.3 & 73 & 1000 & $1.7$ & - & - & - \\
V404~Cyg$^a$ & 3.5 & 11.7 & 7.0 & 56 & 5000 & 18 & - & - & - \\ \hline
\end{tabular}
\\$^1$ The columns are: source name, distance, BH mass, Eddington scaled
luminosity in the 0.5--10~keV band, binary inclination, 
RIAF/disc transition radius, accretion rate (in Eddington units) at $r=r_{tr}$,
jet inclination, outflow rate (in Eddington units), and jet kinetic 
power $\Gamma_{j}(\Gamma_{j}-1)\dot{M}_{jet}c^2$
(normalized to accretion power at $r=10 R_S$).  \\
$^a$ Pure accretion modeling. \\
$^b$ McClintock \& Remillard 2004, and references therein. \\
$^c$ Hjellming \& Rupen 1995. \\
$^d$ The jet angle is assumed to be equal to the inclination angle of the 
binary orbit.
 
\end{table}

\clearpage

\begin{figure}
\epsfig{figure=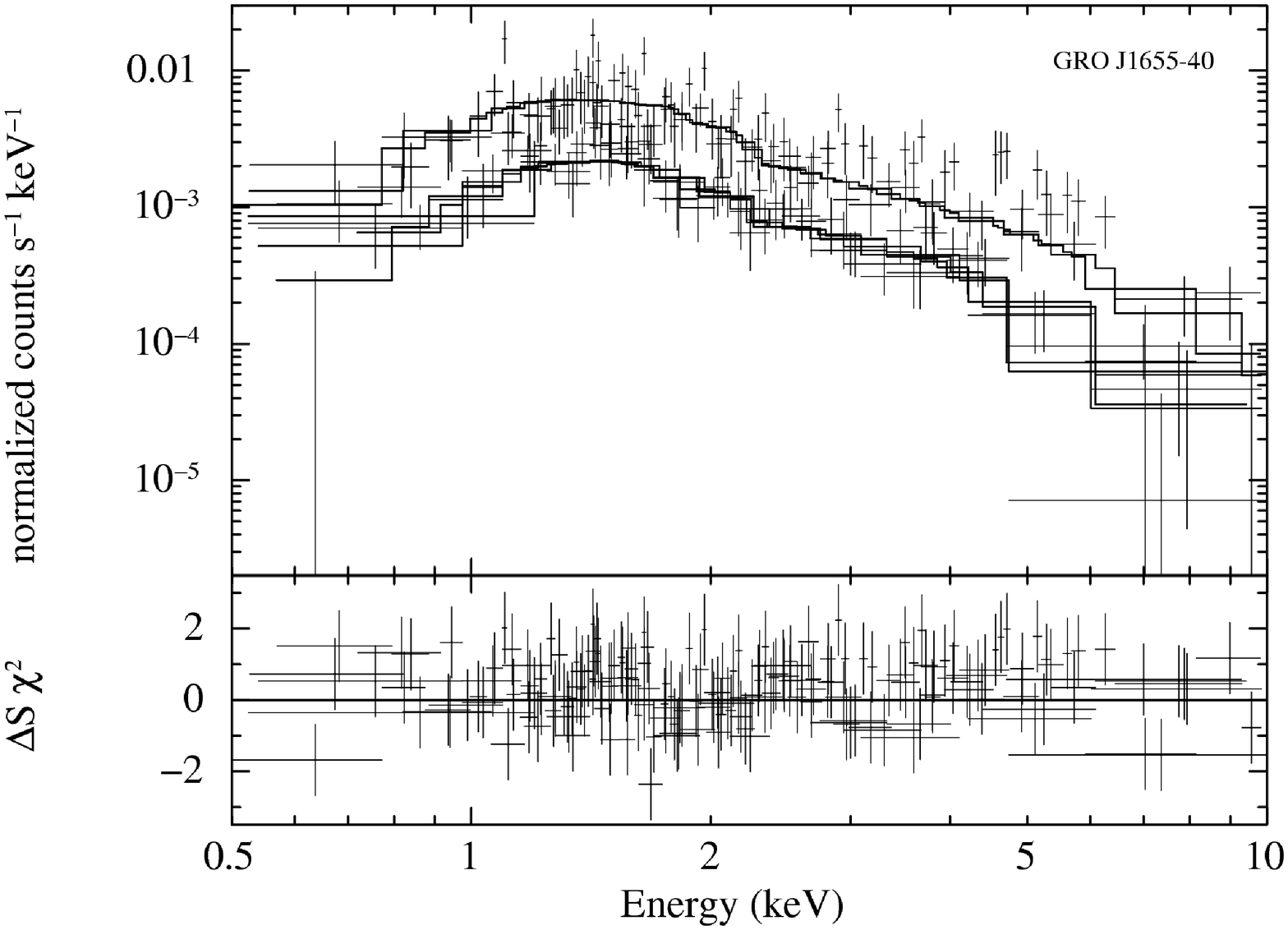,width=8.5cm,angle=0}
\epsfig{figure=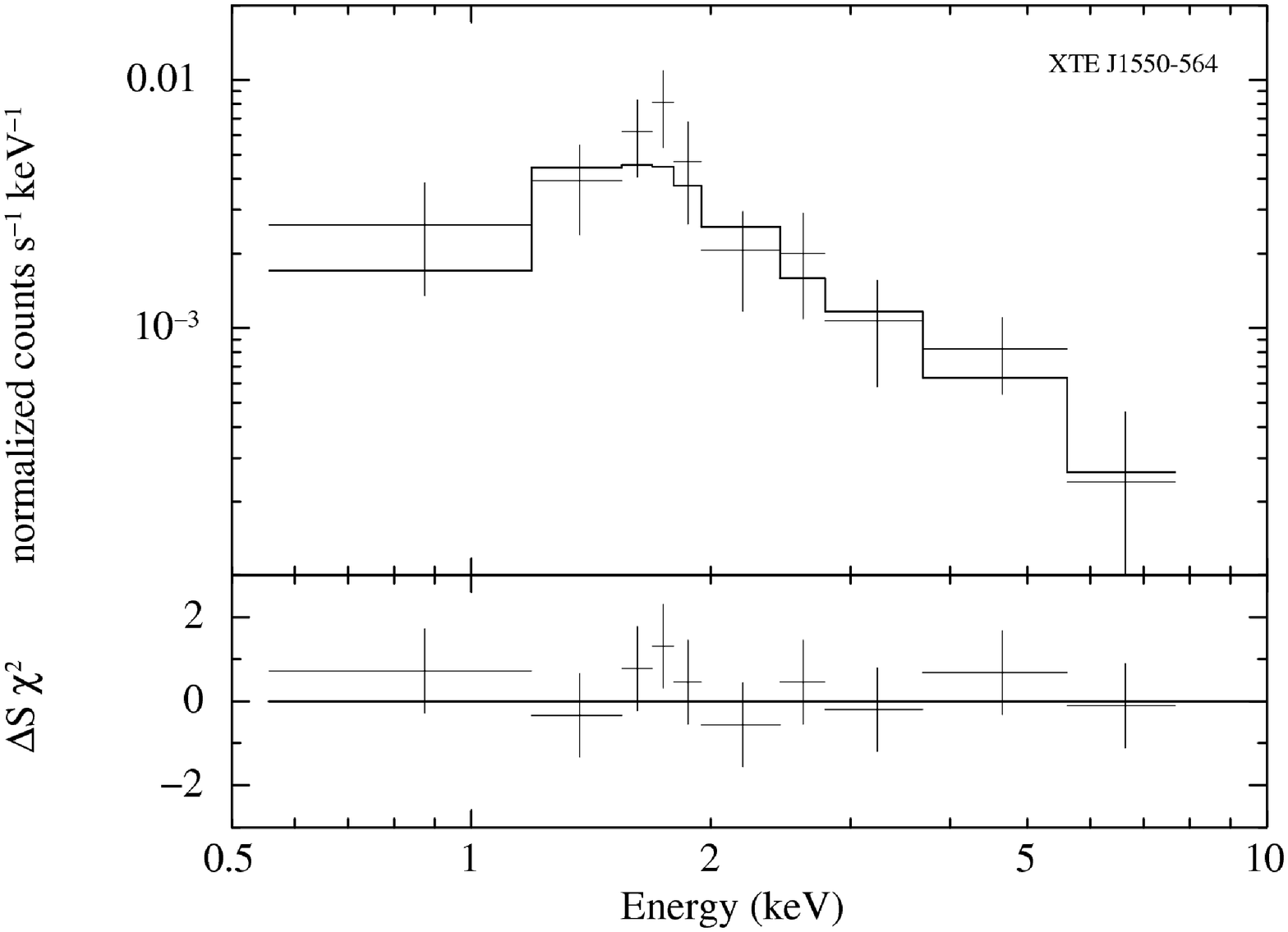,width=8.5cm,angle=0}
\epsfig{figure=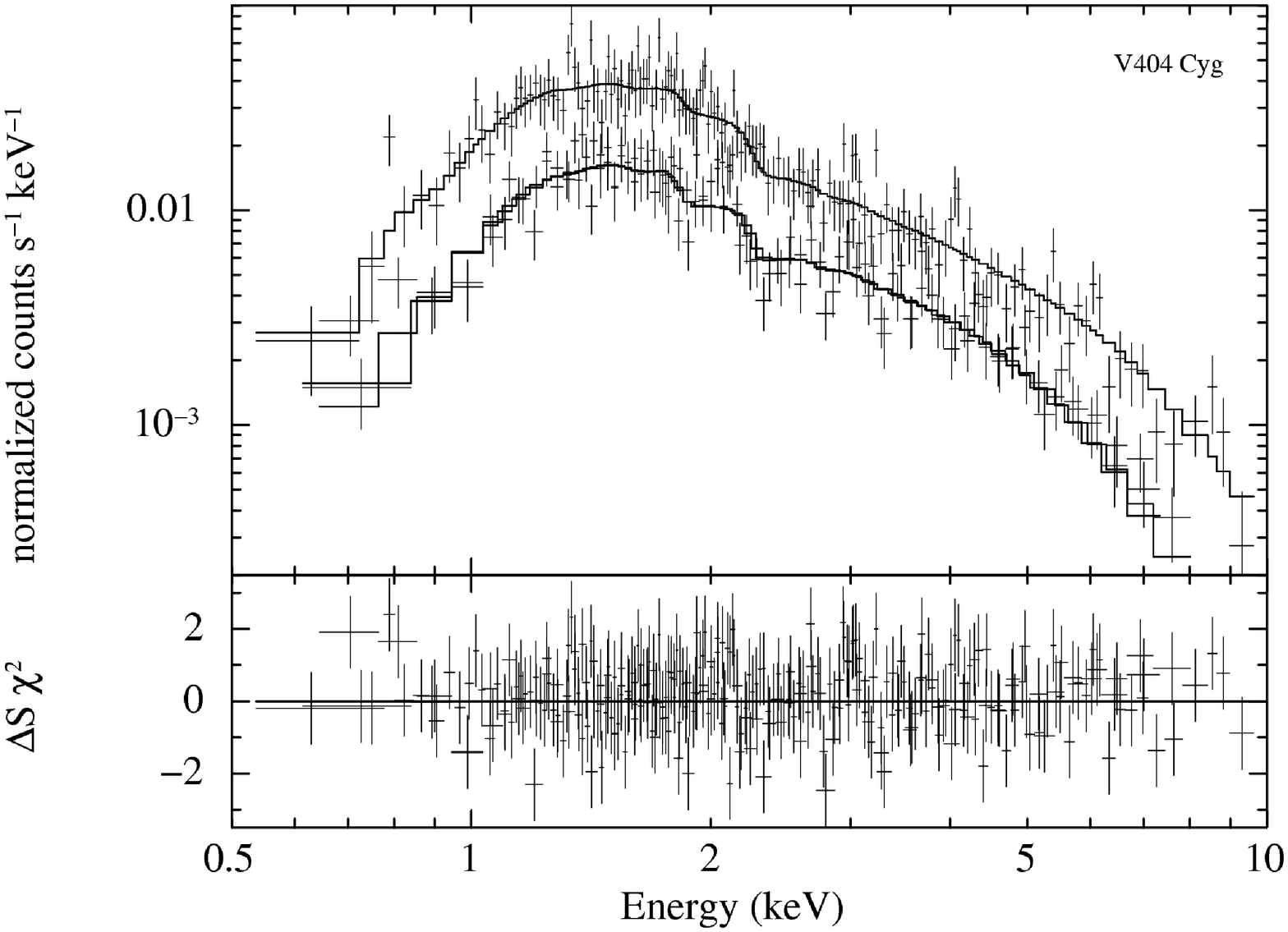,width=8.5cm,angle=0}
\caption{Power-law fit for GRO~J1655$-$40 (top), XTE~J1550$-$564
(middle), and V404~Cyg (bottom). }
\end{figure}

\begin{figure}
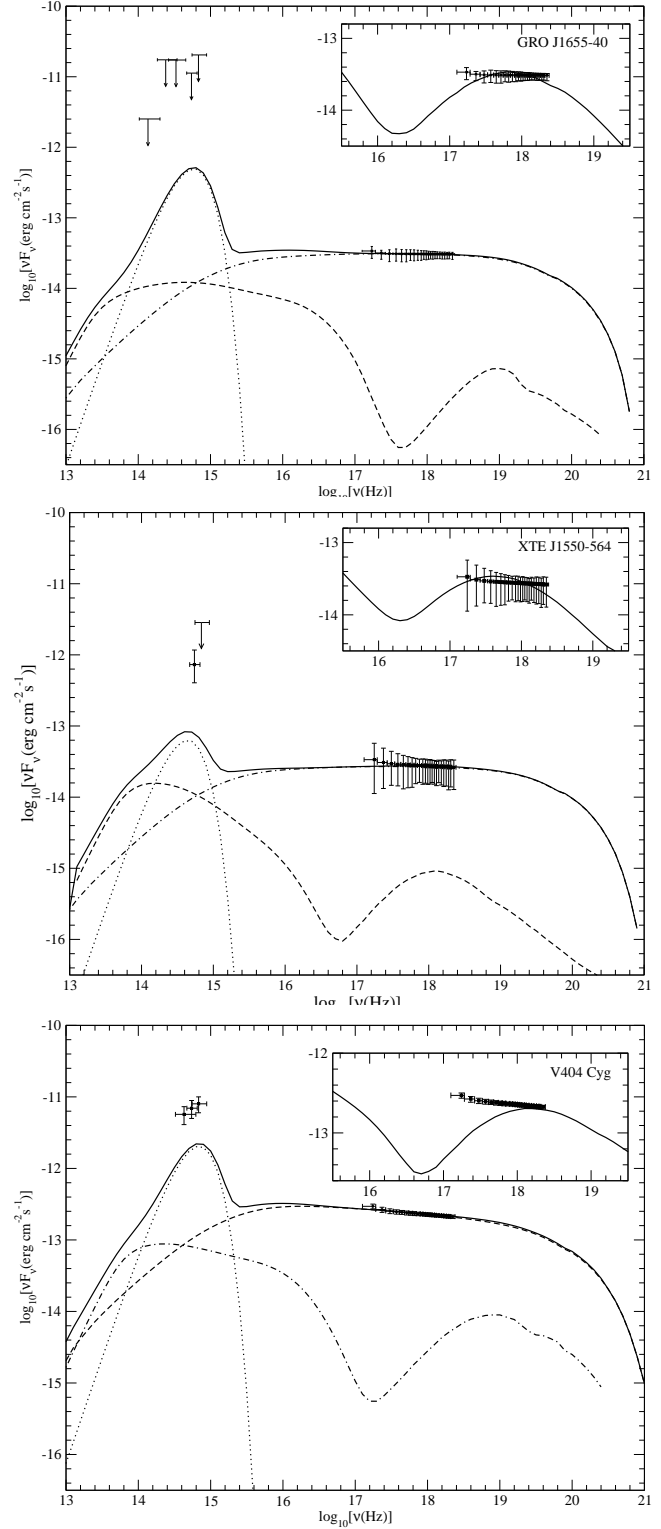

\epsfig{figure=f2a.eps,width=8.5cm,angle=0}
\epsfig{figure=f2b.eps,width=8.5cm,angle=0}
\epsfig{figure=f2c.eps,width=8.5cm,angle=0}
\caption{Spectral energy distribution (SED) for GRO~J1655$-$40 (top),
XTE~J1550$-$564 (middle), and V404~Cyg (bottom).
The solid lines show representative
fits with the coupled accretion-jet model (see text). The dashed, dot-dashed, 
and dotted lines indicate individual contributions from the hot flows, jets, 
and the cool disc, respectively. 
For the purpose of illustration, the optical measurements from the literature
are included. Since these measurements are not simultaneous with the X-ray
measurements, one must exercise caution in drawing conclusions given the
potential variability of the sources in the quiescent state.
The insets show pure accretion modeling, that is clearly inconsistent
with the SED in case of V404~Cyg and GRO~J1655$-$40. The case is ambiguous for
XTE~J1550$-$564. }
\end{figure}

\end{document}